\documentclass[sn-aps]{sn-jnl}% American Physical Society (APS) Reference Style
\usepackage{graphicx}%
\usepackage{multirow}%
\usepackage{amsmath,amssymb,amsfonts}%
\usepackage{amsthm}%
\usepackage{mathrsfs}%
\usepackage[title]{appendix}%
\usepackage{xcolor}%
\usepackage{textcomp}%
\usepackage{manyfoot}%
\usepackage{booktabs}%
\usepackage{algorithm}%
\usepackage{algorithmicx}%
\usepackage{algpseudocode}%
\usepackage{listings}%
\usepackage{siunitx}
\usepackage{xspace}
\newcommand{\um}{\si{\micro m}\xspace}
\newcommand{\uJ}{\si{\micro J}\xspace}
\begin{document}

\title{\bf Origin of optical nonlinearity in plasmonic semiconductor nanostructures}


\author[1]{\fnm{Andrea} \sur{Rossetti}}
\equalcont{These authors contributed equally to this work.}

\author[2]{\fnm{Huatian} \sur{Hu}}
\equalcont{These authors contributed equally to this work.}

\author[3]{\fnm{Tommaso} 
\sur{Venanzi}}
\equalcont{These authors contributed equally to this work.}

\author[4]{\fnm{Adel} \sur{Bousseksou}}

\author[5]{\fnm{Federico} \sur{De Luca}}

\author[1]{\fnm{Thomas} \sur{Deckert}}

\author[3]{\fnm{Valeria} \sur{Giliberti}}

\author[6]{\fnm{Marialilia} \sur{Pea}}

\author[4]{\fnm{Isabelle} \sur{Sagnes}}

\author[4]{\fnm{Gregoire} \sur{Beaudoin}}

\author[7]{\fnm{Paolo} \sur{Biagioni}}

\author[8]{\fnm{Enrico} \sur{Baù}}

\author[9,10]{\fnm{Stefan A.} \sur{Maier}}

\author[8]{\fnm{Andreas} \sur{Tittl}}

\author[1]{\fnm{Daniele} \sur{Brida}}

\author[4]{\fnm{Raffaele} \sur{Colombelli}}

\author*[3,6,11]{\fnm{Michele} \sur{Ortolani}}
\email{michele.ortolani@uniroma1.it}

\author*[2]{\fnm{Cristian} \sur{Ciracì}}
\email{cristian.ciraci@iit.it}

\affil[1]{\orgdiv{Department of Physics and Materials Science}, \orgname{University of Luxembourg}, \orgaddress{\street{162a avenue de la Faïencerie}, \city{Luxembourg}, \postcode{L-1511}, \country{Luxembourg}}}

\affil[2]{\orgdiv{Center for Biomolecular Nanotechnologies}, \orgname{Istituto Italiano di Tecnologia}, \orgaddress{\street{via Barsanti 14},\city{Arnesano}, \postcode{73010}, \country{Italy}}}

\affil[3]{\orgdiv{Center for Life Nano- and Neuro-Science}, \orgname{Istituto Italiano di Tecnologia}, \orgaddress{\street{Viale Regina Elena 291},\city{Rome}, \postcode{00161}, \country{Italy}}}

\affil[4]{\orgdiv{Centre de Nanosciences et de Nanotechnologies, CNRS UMR 9001}, \orgname{University of Paris-Saclay}, \orgaddress{%\street{}, 
\city{Palaiseau}, \postcode{91120}, \country{France}}}

\affil[5]{\orgdiv{Photonics Initiative, Advanced Science Research Center}, \orgname{City University of New York}, \orgaddress{%\street{}, 
\city{New York}, \postcode{10031}, \state{NY}, \country{USA}}}

\affil[6]{\orgdiv{Istituto di Fotonica e Nanotecnologie}, \orgname{Consiglio Nazionale delle Ricerche}, \orgaddress{\street{Via del Fosso del Cavaliere 100}, \city{Rome}, \postcode{00133}, \country{Italy}}}

\affil[7]{\orgdiv{Physics Department}, \orgname{Politecnico di Milano}, \orgaddress{\street{Piazza Leonardo da Vinci 32}, \city{Milan}, \postcode{20133}, \country{Italy}}}

\affil[8]{\orgdiv{Chair in Hybrid Nanosystems, Nano-Institute Munich, Faculty of Physics}, \orgname{Ludwig-Maximilians-Universtität München}, \orgaddress{\street{Königinstraße 10}, \city{München}, \postcode{80539}, \country{Germany}}}

\affil[9]{\orgdiv{School of Physics and Astronomy}, \orgname{Monash University}, \orgaddress{\street{Wellington Rd}, \city{Clayton VIC}, \postcode{3800}, \country{Australia}}}

\affil[10]{\orgdiv{The Blackett Laboratory, Department of Physics}, \orgname{Imperial College London}, \orgaddress{\street{Wellington Rd}, \city{London}, \postcode{SW72AZ}, \country{United Kingdom}}}

\affil[11]{\orgdiv{Dipartimento di Fisica}, \orgname{Sapienza Universit\`a di Roma}, \orgaddress{\street{Piazzale Aldo Moro 2}, \city{Rome}, \postcode{00185}, \country{Italy}}}


\abstract{
%The abstract serves both as a general introduction to the topic and as a brief, non-technical summary of the main results and their implications. Authors are advised to check the author instructions for the journal they are submitting to for word limits and if structural elements like subheadings, citations, or equations are permitted.
The development of nanoscale nonlinear elements in photonic integrated circuits is hindered by the physical limits to the nonlinear optical response of dielectrics, which requires that the interacting waves propagate in transparent volumes for distances much longer than their wavelength.
Here we present experimental evidence that optical nonlinearities in doped semiconductors are due to free-electron and their efficiency could exceed by several orders of magnitude that of conventional dielectric nonlinearities. 
Our experimental findings are supported by comprehensive computational results based on the  hydrodynamic modeling, which naturally includes nonlocal effects, of the free-electron dynamics in heavily doped semiconductors. 
By studying third-harmonic generation from plasmonic nanoantenna arrays made out of heavily n-doped InGaAs with increasing levels of free-carrier density, we discriminate between hydrodynamic and dielectric nonlinearities. As a result, the value of maximum nonlinear efficiency as well as its spectral location can now be controlled by tuning the doping level. Having employed the common material platform InGaAs/InP that supports integrated waveguides, our findings pave the way for future exploitation of plasmonic nonlinearities in all-semiconductor photonic integrated circuits.}

%%================================%%
%% Sample for structured abstract %%
%%================================%%




\maketitle

Nonlinear optics has been historically dominated by experimental configurations where interacting optical beams propagate for distances much longer than the involved wavelengths in bulk nonlinear optical crystals \cite{boyd2008nonlinear}, in optical fibers \cite{brida2014ultrabroadband}, or in integrated photonic waveguides \cite{liu2022photonic}.
More recently, nonlinear metasurfaces, or nanoantenna arrays of subwavelength thickness \cite{biagioni2012nanoantennas} have been introduced to eliminate phase-matching constraints \cite{10.1021/acsphotonics.1c01356}, leading in the latter case to the emergence of \textit{nonlinear plasmonics} \cite{Kauranen:2012ff,10.1088/2040-8986/aac8ed,Krasavin:2017ki}. 
Plasmonic nanoantennas are often used as a sub-wavelength field concentrators to enhance the interaction between light and nonlinear dielectric systems and molecules \cite{10.1103/PhysRevLett.104.207402,10.1021/nl5038819,Shibanuma:2017kv}. Remarkably, it has been shown that the plasmonic nanostructure itself can also provide a sub-diffraction limit source of optical nonlinearity \cite{Kauranen:2012ff}. However, the fundamental origin for this phenomenon is still an unsettled question.

In fact, one can identify at least two fundamental mechanisms for instantaneous (i.e. faster than an optical cycle) nonlinear optical response of a plasmonic structure. The first is the dielectric nonlinearity due to the anharmonic potential of individual electrons within bulk crystals  that leads to a response that is local in nature. The second mechanism is due to the collective motion of free electrons under an external radiation field. Such nonlocal oscillation is ultimately related to the kinetic energy of the free-electron gas and can be modelled by a set of hydrodynamic equations of motion in analogy with a classical fluid. The distinction between these two distinct origins is crucial since there is a physical limit to the maximum dielectric nonlinearity, related to the form of anharmonic potentials  \cite{Kuzyk} that sets upper bounds for the nonlinear susceptibilities $\chi^{(2)}$ and $\chi^{(3)}$. Such limitation however does not apply to free-electron nonlinearities \cite{boyd2008nonlinear}. Therefore, one may ask whether the free-electron nonlinearity is the dominant effect, and if so, whether it can be harnessed to exceed the limited efficiency of conventional dielectric nonlinearity. 

In noble metals, and especially in nanoantenna systems, it is extremely difficult to separately investigate the two contributions, as they coexist in the same volumes, without the possibility of discriminating between the two sources of nonlinearity. Doped semiconductors, on the other hand, offer the possibility of controlling the carrier density via external doping, or via a field-effect gate \cite{Luca.2021,10.1103/physrevlett.129.123902,10.1051/epjam/2022011} thus allowing to tune the free-electron response with respect to the dielectric one. Moreover, semiconductors have much lower free carrier densities ($n_0\sim10^{19}$~cm$^{-3}$) and smaller effective carrier masses $m^* \sim 0.1 m_e$ than noble metals, which results in stronger nonlocal effects and larger nonlinear active volumes $V=l^3\sim(v_{\rm F}/\omega_{\rm p})^3 \propto 1/(n_0^{1/2}m^{*3/2})$, which are $\approx 10^3$ times larger than for noble metals, where $v_{\rm F}$ is the Fermi velocity and $\omega_{\rm p}$ is the plasma frequency \cite{Maack:2017fn,Dias:2018bp}. 

In this work, we combine experiments and theoretical modelling to demonstrate that the predominant contribution to the nonlinear optical response of heavily-doped semiconductor nanoantennas arises from the free electrons.
By measuring third harmonic generation (THG) from nanoantennas with different free-electron density and comparing the obtained efficiency to hydrodynamic model calculations, we unveil that the nonlocal free-electron interaction is the fundamental mechanism of nonlinear plasmonics in doped semiconductors in general.
The material platform employed for the experiment is InGaAs/InP because of its broad appeal for the future exploitation of plasmonic nonlinearities in all-semiconductor photonic integrated circuits (PICs) \cite{10.1364/optica.6.001023,10.1038/s41467-023-44628-7,10.1021/acsphotonics.1c01767}.

%\section*{Theory}
%\textbf{Theoretical framework.}
The fermionic nature of electron-electron interactions manifests itself as an internal pressure in the electron gas that resists the compression induced by an external electromagnetic field.
The effects of such pressure are most apparent near the material surfaces where strong gradients of the carrier concentration and of the electric field occur. 
Free-electron nonlinearities \cite{Krasavin:2017ki} are therefore intrinsically nonlocal, in the sense that the induced currents depend not only on the value of the electric field at a given point but also, through their spatial derivatives, on the value of the fields in the surrounding area \cite{Scalora:2020dq,Maack:2017fn,RodriguezSune:2020ih}.
 %This fundamental equation can be related to the non-interacting Schrödinger equation \cite{Ciraci:2017bp} by introducing the macroscopic charge and current densities, $n({\bf r},t)=\sum_{i\in \rm occ}\phi_i^2$  and  ${\bf J}({\bf r},t)=-\frac{e\hbar}{m}\sum_{i\in \rm occ}\phi_i^2\nabla\chi_i$, respectively, with $\varphi_i=\phi_i e^{j\chi_i}$  being the $i-$th complex orbital, $e$ the electron charge in absolute value, $\hbar$ the reduced Planck constant and $m^*$ the electron mass.
%Within the quasi-classical formalism of the hydrodynamic model , the dynamics of the electron fluid can be described through the macroscopic variables $n(\mathbf{r},t)$ and the $\mathbf{J}(\mathbf{r},t)= -en\mathbf{v}$, the charge density and the current density, respectively ($\mathbf{v}$  being the electron velocity field and $e$ the elementary charge in absolute value).
The many-body nonlinear and nonlocal dynamics of a free-electron fluid under external electric and magnetic fields, $\mathbf{E}(\mathbf{r},t)$ and $\mathbf{H}(\mathbf{r},t)$, is described by the following equation for the electron $n({\bf r},t)$ and current ${\bf J}({\bf r},t)$ densities \cite{Toscano:2015iw, Yan:2015ff, deCeglia:2018ip, Raza:2015ef}:

\begin{equation}
\label{eqn:QHT}
m^*\left(\frac{\partial}{\partial t} -\frac{\mathbf{J}}{en}\cdot \nabla + \gamma\right)\frac{\mathbf{J}}{en} =e\mathbf{E} -\frac{\mathbf{J}}{n} \times \mu_0\mathbf{H} +\nabla \frac{\delta G[n]}{\delta n},
\end{equation}
where $\mu_0$ is the magnetic permeability of vacuum, $e$ is the electron charge, $\gamma$ is the damping rate of free carrier motion and $m^*$ accounts for the band structure in a real solid. The last term on the right-hand side contains the gradient of the functional derivative of the free-energy functional $G[n]$, i.e., the \textit{quantum pressure} in the electron fluid \cite{Ciraci:2016il,Ciraci:2017bp}, which can also be obtained from the Thomas-Fermi screening \cite{Luca.2021}.
Here we are neglecting the spatial dependence of $n$ (spill-out effect), since the systems we will investigate are relatively large ($\sim1$~\um).
Following a perturbation approach, it is possible to derive all nonlinear source terms, see details in Supplementary Information (SI). Here, we report the two THG source terms that appear in the propagating-wave solution of the Maxwell's equations and Eq.~\eqref{eqn:QHT}: the third-order dielectric polarization $\mathbf{P}_{\rm d}^{(3)}$, and the hydrodynamic contribution $\mathbf{S}^{(3)}$ given by the sum of convective and quantum pressure terms:

\begin{subequations}
\label{eqn:THG}
\begin{equation}
\label{eqn:P_THG}
\mathbf{P}_{{\rm d,}3\omega}^{(3)}=\varepsilon_0\chi^{(3)}(\mathbf{E}_\omega\cdot\mathbf{E}_\omega)\mathbf{E}_\omega,   
\end{equation}
\begin{align}
%\begin{equation}
\label{eqn:S_THG}
\mathbf{S}_{3\omega}^{(3)}=&-\frac{\omega^2}{e^2n_0^2}\Big[\nabla  \cdot \mathbf{P}_\omega( \mathbf{P}_\omega\nabla  \cdot \mathbf{P}_\omega+\mathbf{P}_\omega \cdot \nabla \mathbf{P}_\omega )\nonumber\\
&+\mathbf{P}_\omega \cdot \mathbf{P}_1\nabla \nabla  \cdot \mathbf{P}_\omega\Big]+\frac{1}{{27}}\frac{{{\beta ^2}}}{{{e^2}{n_0}^2}}\nabla (\nabla  \cdot {{\mathbf{P}}_\omega})^3%\nonumber\\
%&
%-\frac{1}{4}\frac{c_{\rm vW}}{m^*n_0^2}\nabla\left[ {\nabla (\nabla\cdot{\bf P}_1 )\cdot \nabla  (\nabla\cdot{\bf P}_1)}\nabla\cdot{\bf P}_1 +{\nabla(\nabla\cdot{\bf P}_1)} (\nabla\cdot{\bf P}_1)^2\right]
%\end{equation}
\end{align}
\end{subequations}
\noindent  where ${\bf E}_\omega$ and ${\bf P}_\omega=-{\bf J}_\omega/i\omega$ are the field and total polarization vectors at the fundamental frequency $\omega$ and $\beta=\sqrt{\frac{3}{5}}v_{\rm F}$. One immediately sees that an \textit{effective hydrodynamic susceptibility} cannot be rigorously defined: the additional source  ${\bf S}^{(3)}_{3\omega}$ contains the gradient and the divergence of ${\bf P}_\omega$ and therefore it is strictly nonlocal and proportional to $1/{n_0}^2$. Paradoxically, in noble metals, the high concentration of free-carrier leads to weaker nonlinear contributions.
In the absence of spatial variations of ${\bf P}_\omega$, the hydrodynamic model reduces to the Drude model with the permittivity $\varepsilon_{\rm r}=\varepsilon_\infty\left(1-\frac{\omega_{\rm p}^2}{\omega^2+i\gamma\omega}\right)$, with the plasma frequency $\omega_{\rm p} = \sqrt{ n_0 e^2 / \varepsilon_0\varepsilon_\infty m^*}$, where $\varepsilon_\infty$ is the infinity dielectric constant of the semiconductor (see SI). The plasma wavelength $\lambda_{\rm p} = \frac{2\pi c}{\omega_{\rm p}}$ will be used throughout this article, with $c$ being the speed of light in vacuum.

%\textbf{Experiment.}
Heavily doped semiconductors display a broad range of $\lambda_{\rm p}$ values in the infrared (IR) spectrum depending on $\varepsilon_\infty$,  $m^*$ and $n_0$. In practice, if one restricts to small $m^*$ materials compatible with modern PIC nanofabrication processes, the choice reduces to Ge or SiGe grown on Si (group-IV)\cite{10.1021/acsphotonics.8b00438}, In$_{0.53}$Ga$_{0.47}$As grown on InP (InGaAs/InP), and InAs$_{0.9}$Sb$_{0.1}$ grown on GaSb (III-V) \cite{Taliercio:2019ib}. 
All these material systems allow for a limited dopant incorporation which in practice bounds $\lambda_{\rm p} > 5$~\um , i. e., to the mid-IR.
In this work, we have used electron-doped InGaAs/InP with various dopant densities ($m^* = 0.041 m_e$, $\varepsilon_\infty = 12 \varepsilon_0$, $n_0 \leq 1\times10^{19}$ cm$^{-3}$), and utilized fundamental fields (FF) with center wavelength $\lambda_{\rm FF}$ ranging from 12 to 6~\um to drive the nonlinearity. 

\begin{figure}%
\centering
\includegraphics[width=\textwidth]{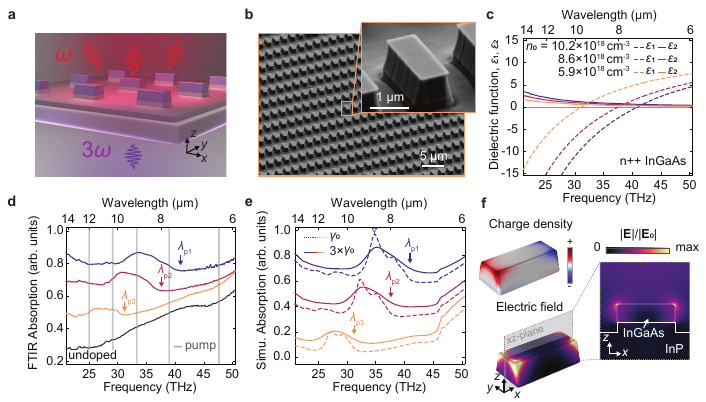}
\caption{$|$ \textbf{Plasmonic antenna arrays.}  \textbf{a,} Schematic illustration of the InGaAs plasmonic antenna on dielectric InP, and the nonlinear experiment producing third harmonic radiations. \textbf{b,} Scanning electron microscope image of the antennas, featuring length, width and thickness approximately equal to 2.2~\um, 0.8~\um and 0.8~\um, respectively. \textbf{c,} Real and imaginary parts of the dielectric function of n-doped InGaAs thin layers used to fabricate the plasmonic nanoantenna.  Experimental \textbf{d}, and theoretical  \textbf{e}  absorption infrared spectra of the antenna arrays for four different doping: undoped (black) to maximum doping (blue). Their corresponding plasma wavelengths were indicated by arrows labeled $\lambda_{\mathrm{p}1,2,3}$. The grey shades indicate the fundamental field wavelengths used in the nonlinear experiments. The dashed and solid lines in simulation \textbf{e} correspond to the results with the original $\gamma_0$ from Drude-model fitting of pristine InGaAs on InP, and broadened decay $3\gamma_0$ which match the experimental absorption, respectively. The additional damping might originate from the geometry imperfections and from the RIE processes, for example due to the inhomogeneous free carrier depletion at the different antenna surfaces. \textbf{f,} the field and induced charge distributions of the main plasmonic resonance marked in \textbf{e}.}\label{fig1}
\end{figure}

Three InGaAs films with different doping levels were grown on undoped InP substrates (see Methods). The InGaAs permittivity $\varepsilon_{\rm r}(\omega)=\varepsilon_1+i\varepsilon_2$ was retrieved from the absolute thick-film ($\sim$2~\um) reflectivity measured by Fourier-transform infrared spectroscopy (FTIR) using the Kramers-Kroenig relations (Fig.~\ref{fig1}c).
A Drude fit to $\varepsilon_{\rm r}(\omega)$ provides $n_0=1.02\times10^{19}$, $8.6\times10^{18}$ and $5.9\times10^{18}$~cm$^{-3}$, for samples 1, 2, and 3 respectively, and a doping-independent scattering rate $\gamma_0=8.9$ THz. $\lambda_{\rm {p}}=9.62$, $7.97$ and $7.33$ \um were directly  obtained as the zero-crossing point of $\varepsilon_1$ in Fig.~\ref{fig1}c \cite{frigerio2016tunability}. 
The InGaAs nanoantennas, consisting of periodically arranged rectangular rods with slightly trapezoidal cross-sections, have been fabricated by etching the excess InGaAs down to the InP substrate by deep reactive ion etching (RIE) through a mask produced by electron-beam lithography, as shown in Fig.~\ref{fig1}b.
The antenna arrays have been characterized by FTIR transmission/reflection microscopy: for light polarized along the antenna axis, they display localized surface plasmon resonances (LSPR) around 8.7, 9.4 and 10.9~\um for the different doping levels, as shown in Fig.~\ref{fig1}d.
These resonances could be well reproduced by numerical full-wave electromagnetic simulations carried out using the finite-element method, as shown in Fig.~\ref{fig1}e (see Methods and SI). The LSPRs are pinned to $\lambda_{\textrm {p}}$ since the antenna length is subwavelength. Fig.\ref{fig1}f displays the induced charge density and electric field of the LSPR, revealing a high-order plasmonic behavior \cite{10.1021/nl200135r}.

\begin{figure}%
\centering
\includegraphics[width=1\textwidth]{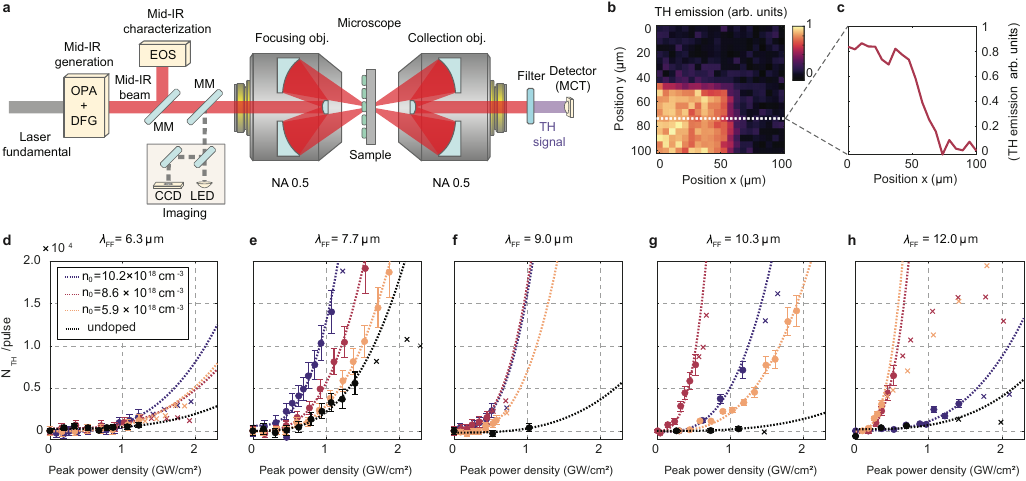}
\caption{$|$ \textbf{Mid-infrared THG from antenna arrays.}  \textbf{a,} Schematic representation of the experimental setup used for the THG experiment. The fundamental beam out of a Yb:KGW laser amplifier is used to drive a tunable mid-IR source, based on difference frequency generation (DFG) between a noncollinear optical parametric amplifier (NOPA) and the laser fundamental. Using a mirror on a magnetic mount (MM) the generated mid-IR transients can be characterized by means of electro-optic sampling (EOS). The mid-IR pulses are coupled into the InGaAs nanoantennas through a reflective microscope and filtered after the interaction with the sample. The emitted third-harmonic signal is collected and measured by an MCT detector. \textbf{b,}	Map of third harmonic emission from one of the nanoantenna arrays used in this experiment. \textbf{c,} Profile of third harmonic emission as the beam position is scanned across the array edge (dotted lines in Fig.~\ref{fig2}b). \textbf{d-h,} Fluence dependence of the THG at different doping levels (color coded) and different fundamental field wavelengths  $\lambda_{\rm FF}$ (different panels). A cubic fit model $N_{\rm TH} = a+\eta_{\rm exp}x^3$ extracts the nonlinear coefficient, omitting data points ($x$) within a saturation regime.}\label{fig2}
\end{figure}

The hydrodynamic nonlinear response shows different regimes depending on the ratio between $\lambda_{\rm FF}$ that drives the emission and $\lambda_{\rm p}$, which in turn depends on $n_0$. To generate different $\lambda_{\rm FF}$ we have employed a pulsed mid-IR laser source, tunable between 5 and 15~\um, and we have tightly focused the beam at the diffraction limit (Fig.~\ref{fig2}a, see Methods). 
The antenna arrays were mounted on a three-dimensional micro-positioner, so as to obtain a two-dimensional map in the focal plane of the third-harmonic emission \cite{10.1038/s41377-018-0108-8}. The strong THG from the antennas is in large contrast to the weak contribution from the substrate as depicted in Figs.~\ref{fig2}b-c ($\chi^{(3)} = 1.4\times10^6$ pm$^2/$V$^2$ in InGaAs, $\chi^{(3)} = 1.0\times10^6$ pm$^2/$V$^2$ in InP \cite{Shibanuma:2017kv}). 

Third-harmonic emission itself is confirmed by measuring a spectrum of the emitted radiation with a dispersive spectrometer (compare SI) while ensuring that other orders of nonlinearity (mainly second) are filtered.
The intensity of the third-harmonic allows to calculate the number of third harmonic photons $N_{\rm TH}$ emitted per pulse by the antenna array as a function of pump peak power density (Fig.~\ref{fig2}d-h). The coefficient of the cubic fits defines the nonlinear efficiency of the THG process $\eta_{\rm exp}$ for each pair of $\lambda_{\rm FF}$ and $n_0$ values.
Above a certain threshold, we observe a deviation from the expected cubic behaviour due to heating \cite{10.1088/1361-6528/aaf5a7} and/or free electron current saturation effects in high driving fields \cite{10.1038/s41377-018-0108-8}, and the corresponding data points are omitted in the fitting.. The values of $\lambda_{\rm FF} =$ 6.3~\um, 7.7~\um, 9.0~\um, 10.3~\um and 12.0~\um  are above, close to, or below  $\lambda_{\rm {p}}=9.62$, $7.97$, and $7.33$ \um for the three samples. The undoped reference nanoantenna sample, which eliminates free-electrons contributions of nonlinearity, showed very weak THG for all $\lambda_{\rm FF}$. 

\begin{figure}%
\centering
\includegraphics[width=1\textwidth]{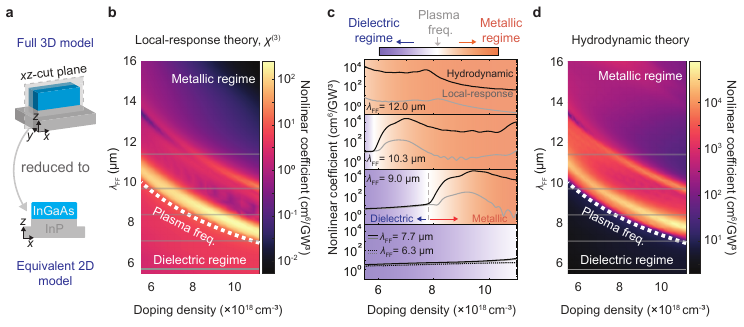}
\caption{$|$ \textbf{Numerical analysis of the equivalent 2D system.}   \textbf{a,} Schematic of the equivalent 2D model. Maps of nonlinear coefficient of the single antenna with different fundamental field wavelengths and doping density based on \textbf{b,} local model with only $\chi^{(3)}$, or \textbf{d,} nonlocal hydrodynamic model with both $\chi^{(3)}$ and hydrodynamic sources. White dotted lines indicate the screened plasma wavelength on different doping densities which infers the condition $Re(\epsilon)=0$. The grey shades indicate the fundamental field wavelengths used in the experiments but with a 0.6~\um  blueshift due to the correction between 2D and 3D models as discussed in the SI.  %The green dashed line, marking the half plasma frequency, indicates the cascaded THG due to $\varepsilon$-near-zero condition that only exists in hydrodynamic theory because the lattice $\chi^{(2)}=0$.
\textbf{c} indicates the specific nonlinear coefficients under the experimental configurations based on the local-response (grey lines) or hydrodynamic (black lines) model. In the dielectric regime, theoretical results of $\lambda_{\rm FF}=6.3$ \um (dashed lines) are comparable with $\lambda_{\rm FF}=7.7$ \um. The colormap indicates different regimes. 
}\label{fig3}
\end{figure}

%\textbf{Discussion.}
To reveal the mechanism at the origin of the free-carrier-density dependent THG, we have numerically solved  Eq.~\eqref{eqn:QHT} together with the wave equation, following a perturbative approach using a finite-element method (see SI), with $n_0$ and $\lambda_{\rm FF}$ as free parameters.
Because of the rapid variations of the fields at the semiconductor surface introduced by the hydrodynamic terms, it is computationally very challenging to perform full three-dimensional (3D) calculations of the antenna system \cite{Vidal-Codina.2021}. Here we used the two-dimensional (2D) equivalent model of Fig. \ref{fig3}a to simulate the single antenna.
The 2D model reproduces the main linear spectral characteristics of the 3D system in Fig.~\ref{fig1}d as long as a systematic shift of approximately $\Delta\lambda \simeq - 0.6$~\um is taken into account (see SI). More importantly, the absorption spectra of the 3D antenna array align well with that of the 3D single antenna (SI), revealing a negligible inter-antenna coupling. This is due to the large gap between every two antennas. This fact validates our strategy of independent-antenna approximation with which the nonlinear coefficient could be scaled by the number of antennas involved when compared with experiments. The numerical nonlinear efficiencies of a single antenna are summarized in Fig.~\ref{fig3}b-d, where we show color maps of the nonlinear coefficient $\eta=\frac{N_{\rm TH}}{I_{\rm FF}^3}$ as a function of $\lambda_{\rm FF}$ and $n_0$, with $N_{\rm TH}$ being the THG photon count per pulse and $I_{\rm FF}$ the fundamental field power density in GW/cm$^2$. In Fig.~\ref{fig3}b we considered a local-response theory with purely dielectric $\chi^{(3)}$ as the only source of nonlinearity (Eq. (\ref{eqn:P_THG})), while in Fig.~\ref{fig3}d the hydrodynamic nonlinear contributions of Eq.~\eqref{eqn:S_THG} were further included apart from the dielectric $\chi^{(3)}$ (see also pure hydrodynamic contributions in SI). The strong dependence of $\eta$ on $n_0$ is markedly different at each $\lambda_{\rm FF}$ as we highlight in Fig. \ref{fig3}c, where we plot a few selected horizontal cuts of the color maps. 

We can identify three different regimes: i) the \textit{dielectric regime}, when $\lambda_{\rm FF} < \lambda_{\rm p}$, i.e., below the dotted line in Figs.~\ref{fig3}b,d and in the bottom panel ($\lambda_{\rm FF}=6.3$ and $7.7$~\um) of Fig.\ref{fig3}c; ii) the \textit{plasmonic resonance regime}, when $\lambda_{\rm FF} \simeq \lambda_{\rm LSPR}$, i. e., the bright regions just above the dotted line in Fig.~\ref{fig3}b,d and in the $\lambda_{\rm FF}=9.0$ and $10.3$~\um panels in Fig.\ref{fig3}c); iii) the \text{metallic regime} when  $\lambda_{\rm FF} > \lambda_{\rm p}$ as in the upper parts of Figs.~\ref{fig3}b,d and in the top panel ($\lambda_{\rm FF}=12.0$~\um) of Fig.\ref{fig3}c.

In the local-response case (grey curves in Fig. \ref{fig3}c) we observe an enhancement of $\eta$ above the dielectric-$\chi^{(3)}$ level of $10^1$ cm$^6 /$ GW$^3$ only in the plasmonic resonance regime (broad peaks at $n_0 \sim 7\times10^{18}$ cm$^{-3}$ for $\lambda_{\rm FF} = 10.3$~\um and $n_0 \sim 9.5\times10^{19}$ for $\lambda_{\rm FF}=9.0$~\um). This is due to the increase of the linear extinction cross-section of the antennas at the LSPR, which effectively increases the polarization field within the material. The peak value of $\eta$ is in the range $10^2$ cm$^6 /$ GW$^3$, 20 times higher than the dielectric-$\chi^{(3)}$ baseline level. In the metallic regime at $\lambda_{\rm FF}=12.0$~\um, $\eta$ even drops to zero for high $n_0$, because there is very small field penetration in the material.

In the hydrodynamic case (black curves in Fig. \ref{fig3}c), $\eta$ is generally much higher than in the local case, apart from the dielectric regime where no plasmons are excited (at nearly all $n_0$ for $\lambda_{\rm FF}=6.3$ and $7.7$~\um and a small range of $n_0 < 8 \times 10^{18}$ cm$^{-3}$ for $\lambda_{\rm FF}=9.0$~\um). In the plasmonic resonance regime, $\eta$ shows a broad enhancement at similar, but not identical, $n_0$ values as for the local theory, and the magnitude of the enhancement is ~200 times stronger (in total, almost 5000 times stronger than the dielectric-$\chi^{(3)}$ baseline level). This cannot be accounted for by the pump extinction enhancement, which affects both source terms equally, therefore it must be due to the hydrodynamic nonlinearity of Eq.~\eqref{eqn:S_THG}. Contrarily to the local model, the nonlinear coefficient enhancement is still visible in the metallic regime, where plasmon fields at the semiconductor surface still exist even out of resonance: the very small field penetration in the bulk does not impact the hydrodynamic nonlinearity, which originates close to the antenna surface, where gradients are strongest. For $\lambda_{\rm FF}=12.0$~\um, $\eta$ is nonzero for high $n_0$ and it is especially strong for decreasing $n_0$, up to $10^4$ cm$^6 /$ GW$^3$. 

\begin{figure}%
\centering
\includegraphics[width=1\textwidth]{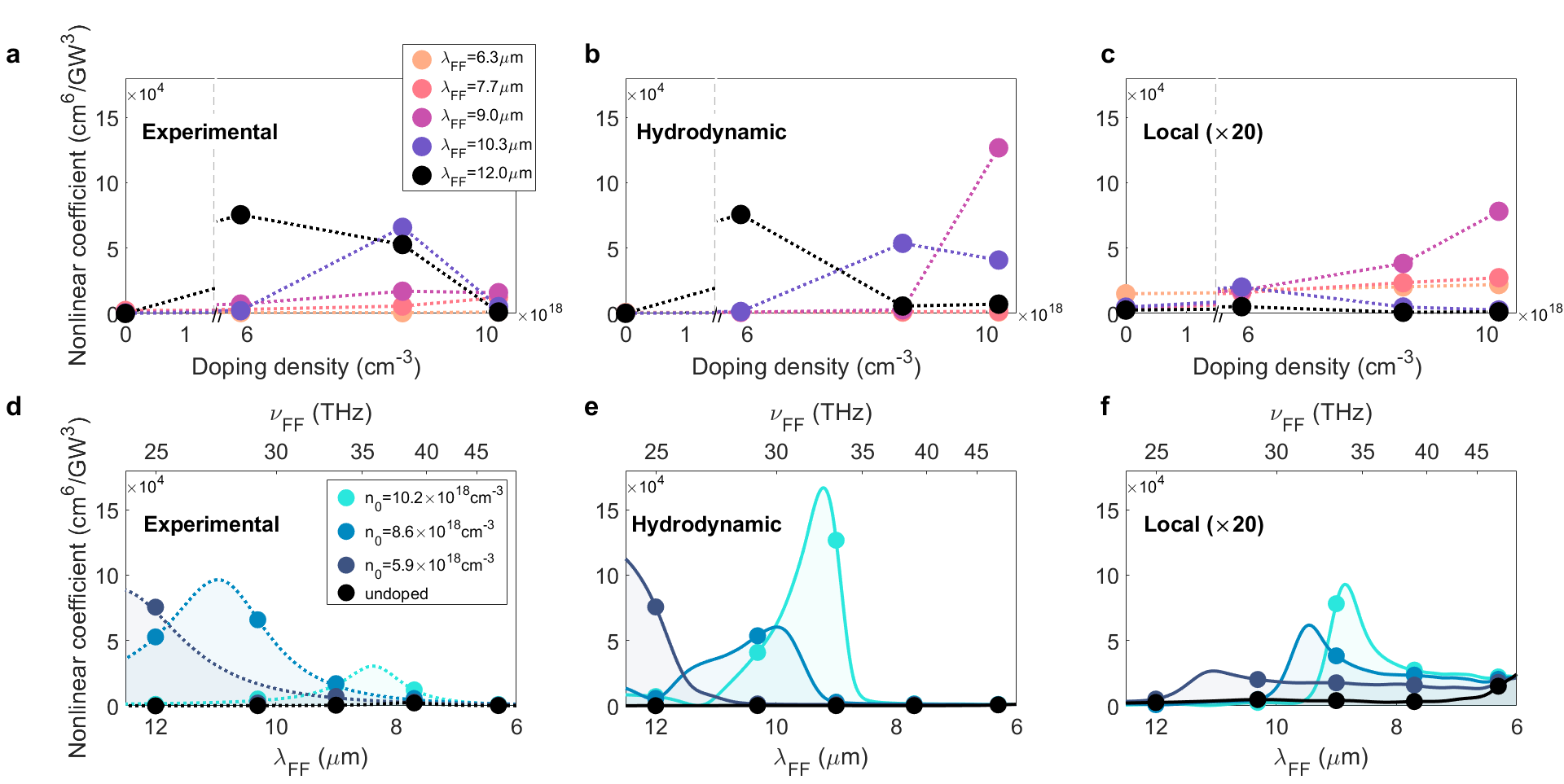}
\caption{$|$ \textbf{Comparison of experiments with different theories.} \textbf{a-c}: Nonlinear coefficients as a function of the doping density for different wavelength $\lambda_{\rm FF}$ of the fundamental field driving the emission. \textbf{a,} Experimental data. Theoretical results from \textbf{b,} hydrodynamic theory results, and \textbf{c,} classical local-response model with only a dielectric $\chi^{(3)}$. \textbf{d-f}: Nonlinear coefficients as a function of $\lambda_{\rm FF}$ acquired from the \textbf{d,} Experiments, \textbf{e,} hydrodynamic theory, and \textbf{f}, classical local-response theory with only a dielectric $\chi^{(3)}$. The experimental data in \textbf{d} is fitted (in the frequency domain) with a Lorentzian curve, here shown as a guide to the eye. Note that the vertical scales of the hydrodynamic model plots match that of the experiments, while those of the local theory plots are 20 times lower. Numerical data were obtained by properly normalizing 2D calculations (see SI).}
%The shades in the theory account for a variation induced by the variation of the time duration of the pulse in the experiments ranging from 360 fs to 430 fs. The nonlinear coefficients predicted by the hydrodynamic theory for 7.7 $\mu \mathrm{m}$ pumping were magnified by 3 times for visualization, whereas the 9.0 $\mu \mathrm{m}$ curve is scaled by 0.3 times.}
\label{fig4}
\end{figure}

We now compare the experimental data with the numerical calculations performed with $\gamma=3\gamma_0$ to account for the broadening of the resonances as observed with linear optical characterization. To compare the numerical THG efficiency with the experimental one, we have estimated the beam width at full-width-half-maximum to be $\sim80$~\um, which implies that $\sim640$ antennas contributed to the measured THG.
The $\eta_{\rm exp}$ retrieved from the cubic fits in Fig.~\ref{fig2}d-h are summarized in Fig.~\ref{fig4}a as a function of $n_0$ and in Fig.~\ref{fig4}d as a function of $\lambda_{\rm FF}$. One can immediately see in Fig.~\ref{fig4}a that three experimental conditions have provided the highest $\eta_{\rm exp}$: $\lambda_{\rm FF} = 12.0$~\um (black dots) and $n_0 =5.9$ and $8.6 \times10^{18}$cm$^{-3}$, and $\lambda_{\rm FF} = 10.3$~\um (violet dots) and $n_0 = 8.6 \times10^{18}$cm$^{-3}$. The experimental results match the predictions from the hydrodynamic model reported in Fig.~\ref{fig4}b, while there is no agreement with the conventional local theory with only dielectric-$\chi^{(3)}$ nonlinearity in Fig.~\ref{fig4}c. Strikingly, both the hydrodynamic and experimental nonlinear coefficients are of the same order of magnitude, $\sim 10^5$ cm$^6 /$ GW$^3$, much higher than the respective values in the local model, $\sim 10^3$ cm$^6 /$ GW$^3$. The experimental data for the sample with highest $n_0 = 10.2 \times10^{18}$cm$^{-3}$ is better analyzed as a function of  $\lambda_{\rm FF}$ in Fig.~\ref{fig4}d: the highest $\eta_{\rm exp}$ for this sample is achieved at $\lambda_{\rm FF} = 7.7$~\um and $9.0$~\um i.e. around its  $\lambda_{\rm LSPR} = 8.7$~\um. Indeed, the position of the efficiency peak around the plasmonic resonance is predicted by both theories and it is approximately recovered by a Lorentzian best-fit of the experimental data for all three samples (dotted curves in Fig.~\ref{fig4}d). Crucially, the experiment agrees well with the main hydrodynamic model prediction in Fig.~\ref{fig4}e, where efficiencies in the metallic regime (long $\lambda_{\rm FF}$) are generally much higher than in the dielectric regime (short $\lambda_{\rm FF}$), while the local model in Fig.~\ref{fig4}f predicts exactly the opposite behaviour.

\textbf{Conclusion}.
The combination of theory and experiments allows to demonstrate that the fundamental origin for THG in optical nanoantennas made of heavily doped semiconductors is the nonlinear collective behaviour of free electrons, described within a hydrodynamic formalism, as opposed to the conventional dielectric nonlinearity described by a local susceptibility $\chi^{(3)}$ independent on the doping level. The experiments show that the efficiency of THG could be up to two orders of magnitude larger than the dielectric susceptibility limit in InGaAs, and this is explained in terms of spatial variation of the amplitude of current and density oscillations, especially in the metallic regime i.e. for pump wavelengths longer than the plasma wavelength, where the local theory predicts a much smaller nonlinear efficiency. 
Although we have highlighted in the main text only third-order terms, which are inversely proportional to the squared carrier density, we have observed in our simulations that cascaded second-order effects equally contribute to the THG efficiency. Second-order contributions contains some terms that do not depend on the inverse charge density and hence would not become negligible in noble metals.  We can thus speculate then that free electrons might also be the predominant source of nonlinearity in all possible plasmonic systems and therefore might be relevant to a wide range of nonlinear experiments that involve gold nanoantenna arrays \cite{10.1021/acs.nanolett.9b02427}. In this context, shorter length scales, stronger fields and higher energy loss might require further developments even beyond  the hydrodynamic description presented, with interesting perspective of understanding collective oscillations in free electron gasses \cite{RodriguezSune:2020ih}.
In addition, the employed semiconductor material platform (InGaAs/InP) is currently under scrutiny to realize photonic integrated circuits in the mid-IR, featuring all-semiconductor waveguides and resonators \cite{10.1364/optica.6.001023,10.1038/s41467-023-44628-7,10.1021/acsphotonics.1c01767}. Plasmonic effects, introduced by selectively doping specific volumes, could provide such photonic integrated circuits with tailored giant nonlinear coefficients. If realized, this new type of tunable nonlinear photonic circuit holds promise for applications ranging from nonlinear signal processing to quantum information distribution. 
Finally, our study underscores the importance of a holistic approach in the design of optical nanoantennas. The local theory allows for the identification of the nonlinear source distribution with the local optical pump intensity patterns. Instead, to quantify the hydrodynamic nonlinearity the full equations of motion of the electron fluid in an external optical field must be solved for each specific geometry. In summary, the nonlocal hydrodynamic response adds a layer of complexity to nonlinear plasmonic device design, but it also unlocks a richer landscape of opportunities.

\section*{Methods}\label{methods}
\textbf{Growth and antenna fabrication.}
The InGaAs thin films were grown by MOCVD (Metal Organic Chemical Vapour Deposition) on InP substrates. The films were doped with Si leading to a n-doping of the material. The thickness of the InGaAs film was about 3$\;\mu$m. The doping levels of the thin films were calculated by measuring the reflectance by means of Fourier-transform infrared spectroscopy (FTIR) and by performing a Drude fit.
The antenna arrays were fabricated by lithography and etching the InGaAs film,  after thinning down the InGaAs epi-layer to 800 nm with chemical etching.

\noindent\textbf{Linear characterization.}
The antenna arrays were investigated by micro-FTIR spectroscopy to measure their plasmonic resonances. The measurements were carried out with a commercial Bruker IFS-66V Michelson interferometer coupled to an infrared microscope (\textit{Hyperion}). The objective was reflective cassegrain-type with a numerical aperture (NA) of 0.4 and a magnification of 15x. The detector is a liquid nitrogen-cooled  Mercury Cadmium Telluride (MCT). The FTIR measurements of the antenna arrays were performed both in reflection ($R$) and in transmission ($T$), and the absorption coefficients shown in Fig 1d were calculated as $1- R - T$.

\noindent\textbf{Nonlinear characterization.}
Our tunable mid-IR source is based on a Yb:KGW laser amplifier, emitting 100~\uJ pulses with 1030~nm central wavelength and operating at a repetition rate of 100~kHz. 
Fundamental wavelength (FW) pulses with energy of 50~\uJ drive a noncollinear optical parametric amplifier (NOPA), delivering broadband near-infrared pulses tunable in the range between 1050 and 1400~nm and 1~\uJ pulse energy. 
The output of the NOPA and the remaining 50~\uJ of laser FW are collinearly focused onto a 1.2-mm-thick GaSe crystal, where p-polarized mid-IR pulses (with pulse energy up to 100~nJ) are generated by means of difference frequency generation (DFG) in a type-II configuration.
The spectrum of the mid-IR pulses can be tuned by suitable selection of the NOPA output central wavelength along with careful adjustments of the phase-matching conditions. The resulting mid-IR transients are characterized by means of electro-optic sampling, yielding for all the excitation pulses used in this work a temporal duration of 400~fs, a bandwidth of 1.5~THz and peak electric fields up to 10~MV/cm.
The mid-IR pulses are coupled into the InGaAs antennas using a confocal microscope working in transmission geometry and based on gold-coated dispersionless Cassegrain reflective objectives with 0.5 numerical aperture (NA). 

The microscope can also be used in reflective geometry to image the sample and locate the antenna arrays. The emitted third harmonic radiation is measured using a liquid-nitrogen-cooled MCT detector and lock-in readout.  In order to filter the fundamental mid-IR pulse from the third-harmonic emission we have used a 5mm thick sapphire window, which acts as a short pass filter with transmission edge at 5~\um. 
In order to filter spurious second harmonic emission from the substrate we have employed either crystalline filters or a monochromator, depending on the excitation wavelength. The monochromator has also been used to record the spectrum of the third-harmonic emission from the antennas.

\noindent\textbf{Simulations.}
We used the finite-element method  (COMSOL Multiphysics) to solve the differential equation system formed by the free-electrons equation and the electromagnetic wave equation in the frequency domain.  The customized coupled equations were implemented using proper weak-form expressions. Overall, three steps, where each step solving for each harmonic ($\omega,2\omega,3\omega$), were used to take both cascaded and direct THG into account, see details in SI.

\backmatter

%\bmhead{Online content}
%Any methods, additional references, Nature Research reporting summaries, source data, extended data, supplementary information, acknowledgements, peer review information; details of author contributions and competing interests; and statements of data and code availability are available at \href{https://doi.org}{https://doi.org}.

\bmhead{Acknowledgments}
D.B., A.R. and T.D. acknowledge funding by the European Research Council (819871) and the European Regional Development Fund (2017-03-022-19). 
S.M. acknowledges the Lee-Lucas Chair in Physics.
A.T. and E.B. acknowledge funding by the Deutsche Forschungsgemeinschaft (DFG, German Research Foundation) under grant numbers EXC 2089/1–390776260 (Germany's Excellence Strategy) and TI 1063/1 (Emmy Noether Program), by the European Research Council (METANEXT, 101078018), and the Center for NanoScience (CeNS).
H.H., T.V., A.B., V.G., M.P., I.S., G.B., E.B., A.T., R.C., M.O., and C.C. acknowledge funding by the European Innovation Council (NEHO, 101046329).

Views and opinions expressed are however those of the authors only and do not necessarily reflect those of the European Union. Neither the European Union nor the granting authority can be held responsible for them.

\end{document}